\documentclass[a4paper,aps,preprintnumbers,showpacs,twocolumn,superscriptaddress,nofootinbib,amsmath,amssymb]{revtex4-1}
\usepackage[dvips]{graphics}
\usepackage{color,hyperref}
\usepackage{times}
\hypersetup{
  colorlinks=true,        
  linkcolor=blue,         
  citecolor=magenta,      
  filecolor=magenta,      
  urlcolor=blue            
}

\def\beq{\begin{equation}}
\def\eeq{\end{equation}}
\def\bear{\begin{eqnarray}}
\def\ear{\end{eqnarray}}

\usepackage{leftidx}
\usepackage{enumitem}
\usepackage{graphicx,subfigure}
\usepackage{dcolumn}
\usepackage{bm}
\usepackage{color}

\begin{document}

\title{Scalar perturbations of nonsingular nonrotating black holes in conformal gravity}

\author{Bobir Toshmatov}
\email{bobir.toshmatov@fpf.slu.cz}
\affiliation{Institute of Physics and Research Centre of Theoretical Physics and Astrophysics, Faculty of Philosophy \& Science, Silesian University in Opava, Bezru\v{c}ovo n\'{a}m\v{e}st\'{i} 13,  CZ-74601 Opava, Czech Republic}
\affiliation{Ulugh Beg Astronomical Institute, Astronomicheskaya 33, Tashkent 100052, Uzbekistan}

\author{Cosimo Bambi}
\email{bambi@fudan.edu.cn}
\affiliation{Center for Field Theory and Particle Physics and Department of Physics, Fudan University, 200433 Shanghai, China}
\affiliation{Theoretical Astrophysics, Eberhard-Karls Universit\"{a}t T\"{u}bingen, 72076 T\"{u}bingen, Germany}

\author{Bobomurat Ahmedov}
\email{ahmedov@astrin.uz}
\affiliation{Ulugh Beg Astronomical Institute, Astronomicheskaya 33, Tashkent 100052, Uzbekistan}
\affiliation{National University of Uzbekistan, Tashkent 100174, Uzbekistan}

\author{Zden\v{e}k Stuchl\'{i}k}
\email{zdenek.stuchlik@fpf.slu.cz}
\affiliation{Institute of Physics and Research Centre of Theoretical Physics and Astrophysics, Faculty of Philosophy \& Science, Silesian University in Opava, Bezru\v{c}ovo n\'{a}m\v{e}st\'{i} 13,  CZ-74601 Opava, Czech Republic}

\author{Jan Schee}
\email{jan.schee@fpf.slu.cz}
\affiliation{Institute of Physics and Research Centre of Theoretical Physics and Astrophysics, Faculty of Philosophy \& Science, Silesian University in Opava, Bezru\v{c}ovo n\'{a}m\v{e}st\'{i} 13,  CZ-74601 Opava, Czech Republic}

\begin{abstract}
We study scalar and electromagnetic perturbations of a family of nonsingular nonrotating black hole spacetimes that are solutions in a large class of conformally invariant theories of gravity. The effective potential for scalar perturbations depends on the exact form of the scaling factor. Electromagnetic perturbations do not feel the scaling factor, and the corresponding quasinormal mode spectrum is the same as in the Schwarzschild metric. We find that these black hole metrics are stable under scalar and electromagnetic perturbations. Assuming that the quasinormal mode spectrum for scalar perturbations is not too different from that for gravitational perturbations, we can expect that the calculation of the quasinormal mode spectrum and the observation with gravitational wave detectors of quasinormal modes from astrophysical black holes can constrain the scaling factor and test these solutions.
\end{abstract}

\maketitle

\section{Introduction}\label{sec-intr}

Despite the successes of Einstein's gravity to explain a large number of observational data~\cite{will}, there have been many attempts to find and study alternative theories of gravity. An important problem in Einstein's gravity is the presence of spacetime singularities in physically relevant solutions. At a singularity, predictability is lost and standard physics breaks down. An attractive possibility to solve the singularity problem is represented by the family of conformally invariant theories of gravity~\cite{cg1,cg2,cg3,cg4,cg5}.

In the context of gravity theories, conformal invariance means that the theory is invariant under a conformal transformation of the metric tensor; that is
\bear
g_{\mu\nu} \rightarrow g^*_{\mu\nu} = \Omega^2 g_{\mu\nu} \, ,
\ear
where $\Omega = \Omega (x)$ is a function of the spacetime coordinates. Einstein's gravity is not conformally invariant. However, it can be made conformally invariant by introducing the auxiliary scalar field $\phi$ (dilaton) as follows:
\bear\label{t1}
\mathcal{L}_1 = \phi^2 R + 6 g^{\mu\nu}
\left( \partial_\mu \phi \right) \left( \partial_\nu \phi \right) \, .
\ear
Another example of the four-dimensional conformal theory of gravity is
\bear\label{t2}
\mathcal{L}_2 = a C_{\mu\nu\rho\sigma} C^{\mu\nu\rho\sigma}
+b R_{\mu\nu\rho\sigma} \tilde{R}^{\mu\nu\rho\sigma} \, .
\ear
In~(\ref{t2}), there is no dilaton field, $C^{\mu\nu\rho\sigma}$ is the Weyl tensor, $R^{\mu\nu\rho\sigma}$ is the Riemann tensor, $\tilde{R}^{\mu\nu\rho\sigma}$ is the dual of the Riemann tensor, and $a$ and $b$ are some constants.

Einstein's gravity is invariant under general coordinate transformations. We say that a metric is singular when its singularities cannot be removed by a coordinate transformation. Otherwise, we speak about ``coordinate singularities'', which are not true singularities of the spacetime, but only an artifact of the coordinate system. Conformally invariant theories of gravity are invariant under both general coordinate transformations and conformal transformations. The key point is that it is always possible to remove a singularity with a suitable family of conformal transformations~\cite{Bambi:1611.00865}. This means that, strictly speaking, conformally invariant theories of gravity are free from spacetime singularities: all observable quantities are independent of the choice of the conformal factor $\Omega$, and the ``singularity'' is not real because it can be removed by a conformal transformation. However, the Universe around us is not conformally invariant. Conformal invariance must be broken, and one possibility is that it is spontaneously broken. In such a situation, Nature must (somehow) select one of the vacua. In the broken phase there are singular and regular metrics, which have different physical properties because there is no conformal symmetry any longer, and we assume that Nature has a mechanism to select a metric in the class of the regular ones. In this paper we are thus interested in the phenomenology of the regular metrics in the broken phase.

Singularity-free, nonrotating and rotating, black hole solutions in conformal gravity were found in~\cite{Bambi:1611.00865,Modesto:2016max}. These solutions are geodetically complete because no massless particle can reach the center of the black hole with a finite value of the affine parameter, and no massive particle can reach the center of the black hole in a finite proper time. These solutions are also singularity-free with respect to the curvature invariants, like the Kretschmann scalar, which are everywhere finite in the spacetime. Note also that conformal invariance is preserved at the quantum level in any finite (i.e. without divergences) quantum field theory of gravity~\cite{Modesto:2014lga}.

In the present paper, we continue the study of the singularity-free black hole metrics of Refs.~\cite{Bambi:1611.00865,Modesto:2016max}. Here, we are interested in possible observational properties that could be exploited to check whether the metric around astrophysical black holes is described by the Kerr solution of Einstein's gravity or by the singularity-free metrics of conformal gravity found in~\cite{Bambi:1611.00865,Modesto:2016max}. There are two approaches to probe the geometry of the strong gravity region of black holes: with electromagnetic radiation~\cite{r-e,r-e2} and with gravitational waves~\cite{r-gw}. In Ref.~\cite{Bambi:1701.00226}, one of us studied the possibility of testing these metrics with the former approach, and in particular with the reflection spectrum of accretion disks. In the present paper, we explore the opportunities offered by gravitational waves.

As a preliminary work, we only study scalar and electromagnetic perturbations. We find that scalar perturbations are sensitive to the exact form of the scaling factor. Assuming that the quasinormal mode spectrum for scalar perturbations is not too different from that for gravitational perturbations\footnote{This assumption is a valid approximation in Einstein's gravity and we expect that the same is true here for a preliminary study such as the present work. The dominant modes are sensitive to the form of the potential, in particular at its maximum. Our potential reduces exactly to that of Schwarzschild for $L=0$, so it should not be too different considering that $L/M < 1.2$, the maximum of the potential is at $r \approx 3M$, and deviations from Schwarzschild enter as $L^2/r^2$ (see next section for more details).}, we can expect to be able to constrain the scaling factor from the observations of quasinormal modes from astrophysical black holes. In the case of electromagnetic perturbations, the equations do not depend on the scaling factor. This is probably related to the conformal symmetry of the Maxwell equations. The result is that the nonsingular black holes and Schwarzschild black holes have the same quasinormal mode spectrum for electromagnetic perturbations.

The content of the paper is as follows. In Section~\ref{sec-spacetime}, we briefly review the singularity-free non-rotating black hole solutions of conformal gravity discussed in~\cite{Bambi:1611.00865,Modesto:2016max}. In Section~\ref{sec-scalar}, we present the key equations for the study of scalar perturbations. In Section~\ref{sec-em}, we do the same for electromagnetic perturbations. In Section~\ref{num-results}, we present the numerical results of scalar perturbations in the singularity-free non-rotating black hole background. Summary and conclusions are reported in Section~\ref{sec-conclusions}. Throughout the paper we use natural units in which $c = G_{\rm N} = 1$ and a metric with signature $(-+++)$.

\section{Nonsingular nonrotating black hole metric in conformal gravity}\label{sec-spacetime}

The line element of the family of singularity-free nonrotating black hole spacetimes in conformal gravity is given by~\cite{Bambi:1611.00865,Modesto:2016max}
\bear\label{metric-Schw}
ds^{\ast2}=S(r)ds_{Schw}^2 \, ,
\ear
where $ds_{Schw}^2$ is the line element of the Schwarzschild metric in Schwarzschild coordinates
\bear\label{Schwarzschild}
ds_{Schw}^2 &=& -f(r)dt^2+ \frac{dr^2}{f(r)}+r^2d\Omega^2 \, , \\
f (r) &=& 1-\frac{2M}{r} \, ,
\ear
and $S(r)$ is the scaling factor
\bear\label{conf1}
S(r)=\left(1+\frac{L^2}{r^2}\right)^{2N} \, .
\ear
In Eq.~(\ref{conf1}), $N=1,2,3,...$ and $L$ is a new scale. $L$ may be expected to be either of order of the Planck scale or of the order of $M$, because these are the only two scales of the system. If $L$ is of the order of the Planck scale, deviations from the Schwarzschild metric are negligible in the case of astrophysical black holes of at least a few Solar masses, and we can unlikely test these metrics. In the rest of the paper we will thus assume that $L$ is of the order of $M$, so that we can hope to be able to distinguish these solutions from the Schwarzschild metric with observations.

The metric in Eq.~(\ref{metric-Schw}) is singularity-free in the sense that it is both geodetically complete and the curvature invariants (e.g. the Kretschmann scalar) are finite everywhere in the spacetime. Geodesic completion is realized because massless particles cannot reach the center for a finite value of the affine parameter and massive particle cannot reach the center in a finite proper time. For more details, see Ref.~\cite{Bambi:1611.00865}. The formation of these singularity-free black holes from gravitational collapse was studied in Ref.~\cite{shila}, while the energy conditions of these spacetimes were studied in Ref.~\cite{Toshmatov:2017kmw}.

\section{Scalar perturbations}\label{sec-scalar}

In this section, we briefly study the scalar perturbations of the regular nonrotating black hole spacetimes in conformal gravity of Eq.~(\ref{metric-Schw}). As pointed out in~\cite{Konoplya-Zhidenko}, if there is no backreaction on the background, the perturbations of black hole spacetimes can be studied not only by adding the perturbation terms into the spacetime metric, but also by inducing fields to the spacetime metric.

\begin{figure*}[th]
\begin{center}
\includegraphics[width=0.32\linewidth]{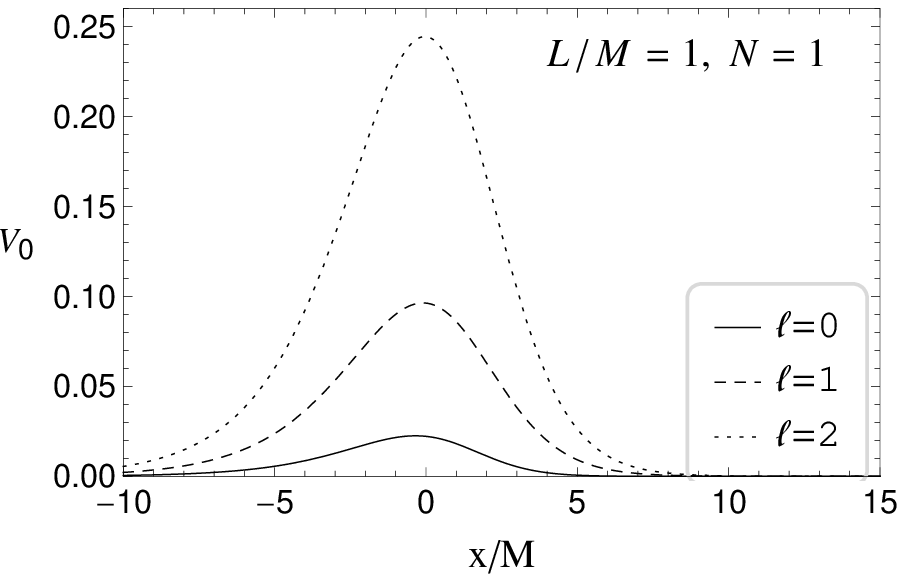}
\includegraphics[width=0.32\linewidth]{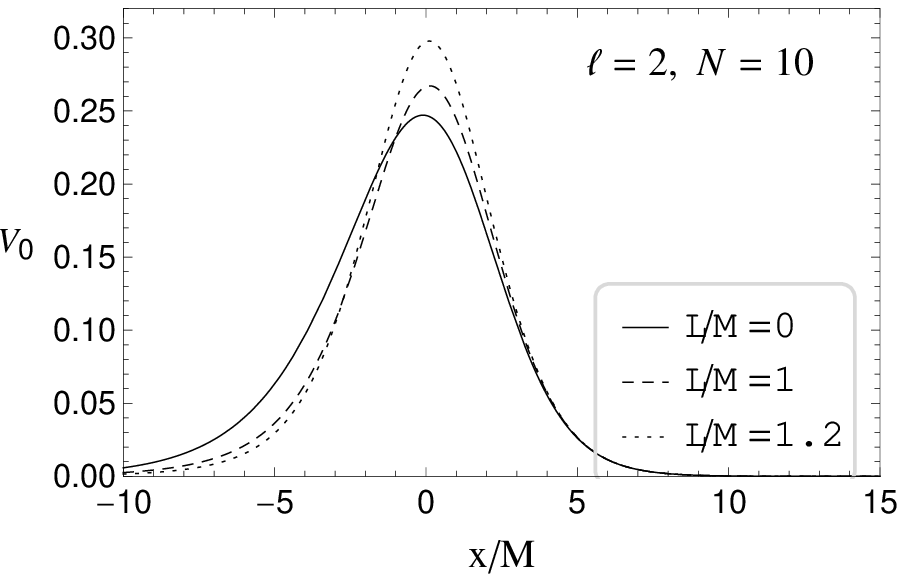}
\includegraphics[width=0.32\linewidth]{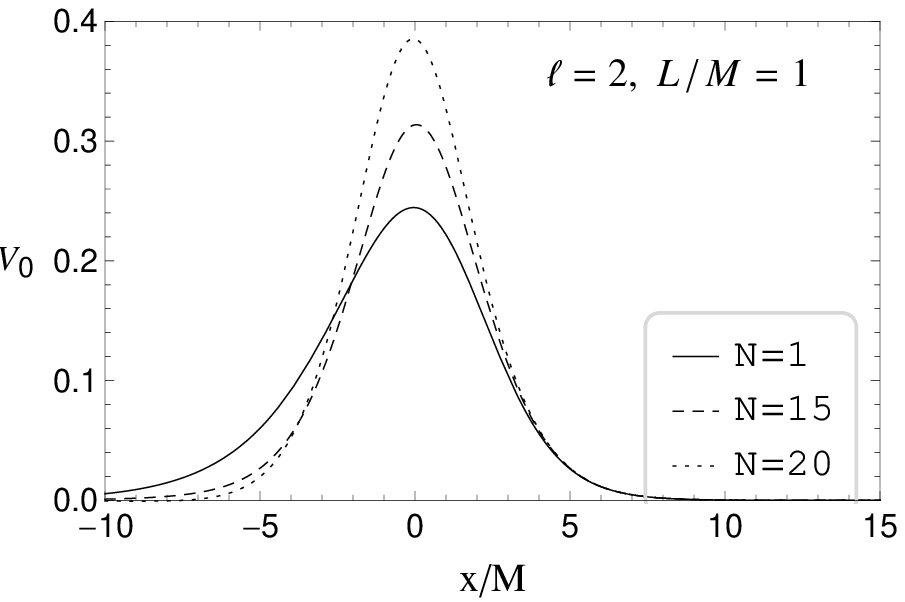}
\end{center}
\caption{\label{fig-potential} Dependence of the effective potential of scalar perturbations on the tortoise coordinate for different values (from left to right) of the angular quantum number $\ell$, conformal factor $L$, and the degree $N$. The values of the parameters are given in the plots.}
\end{figure*}

The equation of motion of the massless scalar field $\Phi$ in curved spacetime is given by the covariant Klein-Gordon equation
\bear\label{K-G}
\frac{1}{\sqrt{-g}}\partial_\mu(g^{\mu\nu}\sqrt{-g}\partial_\nu\Phi)=0\ ,
\ear
where the wave function $\Phi$ is a function of the coordinates $(t,r,\theta,\phi)$. In order to separate the variables in the Klein-Gordon equation, we choose the wave functions as
\bear
\Phi=\frac{1}{r\sqrt{S}}\Psi_0(t,r)Y_\ell^m(\theta,\phi)\ ,
\ear
where $Y_\ell^m(\theta,\phi)$ is the so called spherical harmonic function of degree $\ell$ and order $m$, related to the angular coordinates $\theta$ and $\phi$. We can thus write the wave equation for the scalar field as
\bear\label{wave-eq}
\left(-\frac{\partial^2}{\partial t^2}+\frac{\partial^2}{\partial x^2}+V_0\right)\Psi_0=0\ ,
\ear
where $x$ is the ``tortoise'' (Regge-Wheeler) coordinate. It is related to the radial coordinate by
\bear\label{tortoise}
\frac{\partial}{\partial x}=f\frac{\partial}{\partial r} .
\ear
The potential $V_0$ is defined as
\bear\label{sc-potential}
V_0=f\left[\frac{\ell(\ell+1)}{r^2}+ \frac{1}{r\sqrt{S}}\left(f(r\sqrt{S})'\right)'\right],
\ear
where, here and in what follows, the prime ``$'$'' stands for the derivative with respect to the radial coordinate $r$. The potential~(\ref{sc-potential}) vanishes, $V_0\rightarrow0$, at infinity, $r\rightarrow\infty$, and at the event horizon, $r=2M$. Fig.~\ref{fig-potential} shows the behavior of the effective potential for scalar perturbations for different values of its parameters.
It is well known that the multipole number $\ell$ increases the height of the potential. It is known from the paper~\cite{Bambi:1701.00226} that, if we assume $L \propto M$ and $N = 1$, then astrophysical observations require $L/M < 1.2$. One can see from Fig.~\ref{fig-potential} that, increasing the value of the parameter $L$ in the range of $L/M<1.2$, the height of the potential slightly decreases, while the degree $N$ increases slightly the height of the potential.

\section{Electromagnetic perturbations}\label{sec-em}

In this section we present the electromagnetic perturbations in the regular non-rotating spacetime in conformal gravity of Eq.~(\ref{metric-Schw}). The source-free electromagnetic field in curved spacetime is governed by Maxwell's equations
\bear\label{Maxwell}
\nabla_\nu F^{\mu\nu}=0\ .
\ear
where the electromagnetic tensor $F_{\mu\nu}$ is defined by the electromagnetic potential $A_\mu$ as
\bear\label{em-tensor}
F_{\mu\nu}=\partial_\mu A_\nu-\partial_\nu A_\mu\ ,
\ear
Maxwell's equations in the background~(\ref{metric-Schw}) are written in the form:
\bear\label{Maxwell2}
&&\partial_t F^{\mu t}+\frac{1}{S^2r^2}\partial_r\left(S^2r^2F^{\mu r}\right)\nonumber\\&&+\frac{1}{\sin\theta}\partial_\theta \left(\sin\theta F^{\mu\theta}\right)+ \partial_\phi F^{\mu\phi}=0.
\ear
In spherically symmetric backgrounds, the electromagnetic potential $A_\mu$ can be expanded in 4-dimensional vector spherical harmonics as~\cite{Cardoso-Lemos:PRD:2001,Molina:PRD:2016}
\bear\label{4-potential}
A_\mu(t,r,\theta,\phi)&=&\sum_{\ell,m}\left(\left[
\begin{array}{c}
0 \\
0\\
\frac{a^{\ell m}(t,r)}{\sin\theta}\partial_\phi Y_{\ell m}(\theta,\phi)\\
-a^{\ell m}(t,r)\sin\theta\partial_\theta Y_{\ell m}(\theta,\phi)\\
\end{array}\right]\right.\nonumber\\ &&\left.+\left[
\begin{array}{c}
d^{\ell m}(t,r) Y_{\ell m}(\theta,\phi)\\
h^{\ell m}(t,r) Y_{\ell m}(\theta,\phi)\\
k^{\ell m}(t,r)\partial_\theta Y_{\ell m}(\theta,\phi)\\
k^{\ell m}(t,r)\partial_\phi Y_{\ell m}(\theta,\phi)\\
\end{array}\right]\right),
\ear
where the first and second terms have $(-1)^{\ell+1}$ (axial components of the expansion) and $(-1)^{\ell}$ (polar components the expansion) parities, respectively.

\subsection{Axial electromagnetic perturbations}

The non-vanishing components of the electromagnetic tensor~(\ref{em-tensor}) of the axial components of the 4-potential~(\ref{4-potential}) are given by
\bear
F_{t\theta}&&=\frac{1}{\sin\theta}\partial_t a^{\ell m}\partial_\phi Y_{\ell m}, \nonumber\\ F_{t\phi}&&=-\sin\theta\partial_t a^{\ell m}\partial_\theta Y_{\ell m}, \nonumber\\F_{r\theta}&&=\frac{1}{\sin\theta}\partial_r a^{\ell m}\partial_\phi Y_{\ell m},\\
F_{r\phi}&&=-\sin\theta\partial_r a^{\ell m}\partial_\theta Y_{\ell m},\nonumber\\
F_{\theta\phi}&&=-a^{\ell m}\left[\partial_\theta(\sin\theta\partial_\theta Y_{\ell m})+\frac{1}{\sin\theta}\partial_\phi^2Y_{\ell m}\right]\nonumber\\&&=-\ell(\ell+1)a^{\ell m}\sin\theta.\nonumber
\ear
From the relation $F^{\mu\nu}=g^{\alpha\mu}g^{\beta\nu}F_{\alpha\beta}$, we calculate the non-vanishing components of $F^{\mu\nu}$
\bear\label{non-zero}
&&F^{t\theta}=-\frac{1}{S^2fr^2\sin\theta}\partial_t a^{\ell m}\partial_\phi Y_{\ell m}, \nonumber\\ &&F^{t\phi}=\frac{1}{S^2fr^2\sin\theta}\partial_t a^{\ell m}\partial_\theta Y_{\ell m}, \nonumber\\ &&F^{r\theta}=\frac{f}{S^2r^2\sin\theta}\partial_r a^{\ell m}\partial_\phi Y_{\ell m},\\ &&F^{r\phi}=-\frac{f}{S^2r^2\sin\theta}\partial_r a^{\ell m}\partial_\theta Y_{\ell m}, \nonumber\\&&F^{\theta\phi}=\frac{\ell(\ell+1)}{S^2r^4\sin\theta}a^{\ell m} Y_{\ell m}.\nonumber
\ear
Now, plugging the expressions in~(\ref{non-zero}) into Eq.~(\ref{Maxwell2}), for the cases of $\mu=\theta$ and $\mu=\phi$ we obtain the same equation
\bear
\partial_t^2a^{\ell m}-f\partial_r(f\partial_ra^{\ell m})+f\frac{\ell(\ell+1)}{r^2}a^{\ell m}=0,
\ear
Here, introducing the tortoise coordinates~(\ref{tortoise}) and considering $a^{\ell m}(t,r)=\Psi_{1a}(t,r)$ we obtain the wave equation
\bear
\left(-\frac{\partial^2}{\partial t^2}+\frac{\partial^2}{\partial x^2}+V_{1a}\right)\Psi_{1a}=0\ ,
\ear
with the effective potential for the axial electromagnetic perturbations
\bear\label{axial-potential}
V_{1a}=f\frac{\ell(\ell+1)}{r^2}.
\ear
Note that electromagnetic perturbations do not depend on the scaling factor $S$. This result is valid for any spherically symmetric black hole metric of the form~(\ref{Schwarzschild}), independently of the exact expression of $f$, and for any conformal transformation.

\subsection{Polar electromagnetic perturbations}

The non-vanishing components of the electromagnetic tensor~(\ref{em-tensor}) of the polar components of the 4-potential~(\ref{4-potential}) are given by
\bear
F_{tr}&&=(\partial_t h^{\ell m}-\partial_r d^{\ell m})Y_{\ell m}, \nonumber\\ F_{t\theta}&&=(\partial_t k^{\ell m}-d^{\ell m})\partial_\theta Y_{\ell m}, \nonumber\\F_{t\phi}&&=(\partial_t k^{\ell m}-d^{\ell m})\partial_\phi Y_{\ell m},\\
F_{r\theta}&&=(\partial_r k^{\ell m}-h^{\ell m})\partial_\theta Y_{\ell m},\nonumber\\
F_{r\phi}&&=(\partial_r k^{\ell m}-h^{\ell m})\partial_\phi Y_{\ell m}.\nonumber
\ear
From the relation $F^{\mu\nu}=g^{\alpha\mu}g^{\beta\nu}F_{\alpha\beta}$, we calculate the non-vanishing components of $F^{\mu\nu}$
\bear\label{polar2}
&&F^{tr}=-\frac{1}{S^2}(\partial_t h^{\ell m}-\partial_r d^{\ell m})Y_{\ell m}, \nonumber\\ &&F^{t\theta}=-\frac{1}{S^2fr^2}(\partial_t k^{\ell m}-d^{\ell m})\partial_\theta Y_{\ell m}, \nonumber\\ &&F^{t\phi}=-\frac{1}{S^2fr^2\sin^2\theta}(\partial_t k^{\ell m}-d^{\ell m})\partial_\phi Y_{\ell m},\\ &&F^{r\theta}=\frac{f}{S^2r^2}(\partial_r k^{\ell m}-h^{\ell m})\partial_\theta Y_{\ell m}, \nonumber\\&&F^{r\phi}=\frac{f}{S^2r^2\sin^2\theta}(\partial_r k^{\ell m}-h^{\ell m})\partial_\phi Y_{\ell m}.\nonumber
\ear
Plugging the expressions in~(\ref{polar2}) into Eq.~(\ref{Maxwell2}) we have three independent second order differential equations: for $\mu=t$~\footnote{This equation has a misprint in~\cite{Molina:PRD:2016} that was corrected here.}
\bear\label{dif-eq1}
f\partial_r[r^2(\partial_rd^{\ell m}-\partial_th^{\ell m})]-\ell(\ell+1)(\partial_tk^{\ell m}-d^{\ell m})=0,\nonumber\\
\ear
for $\mu=r$
\bear\label{dif-eq2}
\frac{r^2}{f}\partial_t(\partial_th^{\ell m}-\partial_rd^{\ell m})-\ell(\ell+1)(\partial_rk^{\ell m}-h^{\ell m})=0,\nonumber\\
\ear
and $\mu=\theta$ and $\mu=\phi$ give the same equation
\bear\label{dif-eq3}
f\partial_r[f(h^{\ell m}-\partial_rk^{\ell m})]-\partial_t(d^{\ell m}-\partial_tk^{\ell m})=0,
\ear
Again we have differential equations which are independent from the conformal factor $S$, i.e., Eqs.~(\ref{dif-eq1})-~(\ref{dif-eq3}) correspond to the ones for the any spherically symmetric black hole spacetimes in the form~(\ref{Schwarzschild}). This indicates that polar electromagnetic perturbations also do not depend on the conformal factor $S$.

Since the derivation of the single, the second order differential master equation from Eqs.~(\ref{dif-eq1})-~(\ref{dif-eq3}) has been presented several previous works (for example, see -- Refs.~\cite{Cardoso-Lemos:PRD:2001,Molina:PRD:2016} and references therein.), we just briefly present the ``guidance''. There are several methods of deriving the master equation. For instance~\footnote{Another way of the derivation of the master equations is introducing the new variable
\bear
\partial_r(fh^{\ell m})-\frac{1}{f}\partial_td^{\ell m}=\frac{\ell(\ell+1)}{r^2}k^{\ell m},\nonumber
\ear
and substituting it into Eq.~(\ref{dif-eq3}).}, by introducing the new variable
\bear\label{new-variable}
\partial_th^{\ell m}-\partial_rd^{\ell m}=\frac{\ell(\ell+1)}{r^2}p^{\ell m},
\ear
and substituting~(\ref{new-variable}) into Eqs.~(\ref{dif-eq1}) and~(\ref{dif-eq2}), and differentiating them with respect to $r$ and $t$, respectively, and finally summing up them, one obtains the master equation
\bear
\partial_t^2p^{\ell m}-f\partial_r(f\partial_rp^{\ell m})+f\frac{\ell(\ell+1)}{r^2}p^{\ell m}=0,
\ear
Finally, defining $\Psi_{1p}(t,r)=p^{\ell m}(t,r)$ and introducing the tortoise coordinate~(\ref{tortoise}), one arrives at the equation
\bear\label{master-eq2}
\left(-\frac{\partial^2}{\partial t^2}+\frac{\partial^2}{\partial x^2}+V_{1p}\right)\Psi_{1p}=0\ ,
\ear
with the effective potential for the polar electromagnetic perturbations
\bear\label{polar-potential}
V_{1p}=f\frac{\ell(\ell+1)}{r^2}.
\ear
Thus, the axial and polar modes of electromagnetic perturbations are described by the same effective potentials which are independent on the conformal factor $S$, as presented in~(\ref{axial-potential}) and~(\ref{polar-potential}).
While we do not have any general proof, it is likely that this is due to the conformal symmetry of the vacuum Maxwell equations.

The effective potentials for scalar~(\ref{sc-potential}) and electromagnetic~(\ref{axial-potential}),~(\ref{polar-potential}) perturbations can be written in a compact form as
\bear\label{potentials}
V_s=f\left[\frac{\ell(\ell+1)}{r^2}+ \frac{1-s^2}{r\sqrt{S}}\left(f(r\sqrt{S})'\right)'\right],
\ear
where $s=0$ and $s=1$ represent scalar and electromagnetic perturbations, respectively.

\section{Numerical results}\label{num-results}

\subsection{Evolution of the scalar perturbation}

\begin{figure*}[th]
\begin{center}
\includegraphics[width=0.32\linewidth]{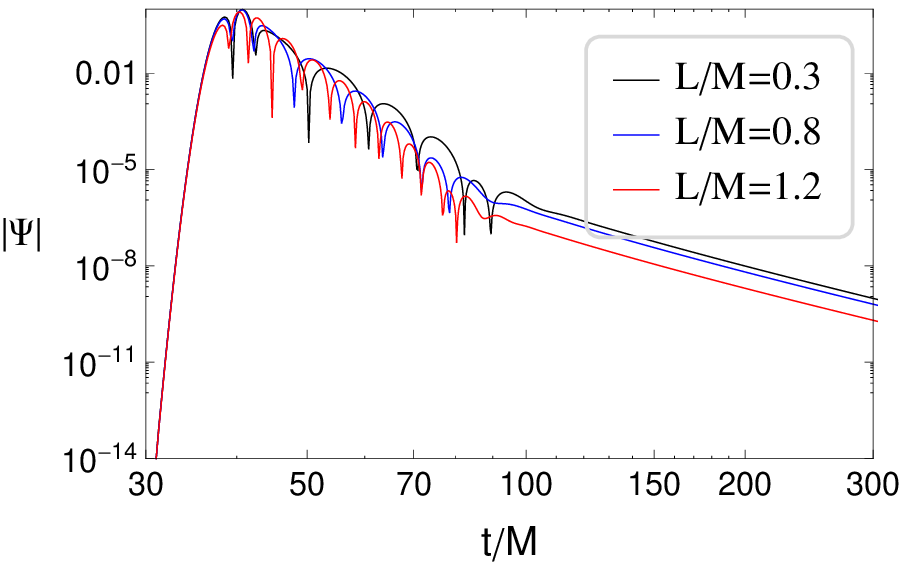}
\includegraphics[width=0.32\linewidth]{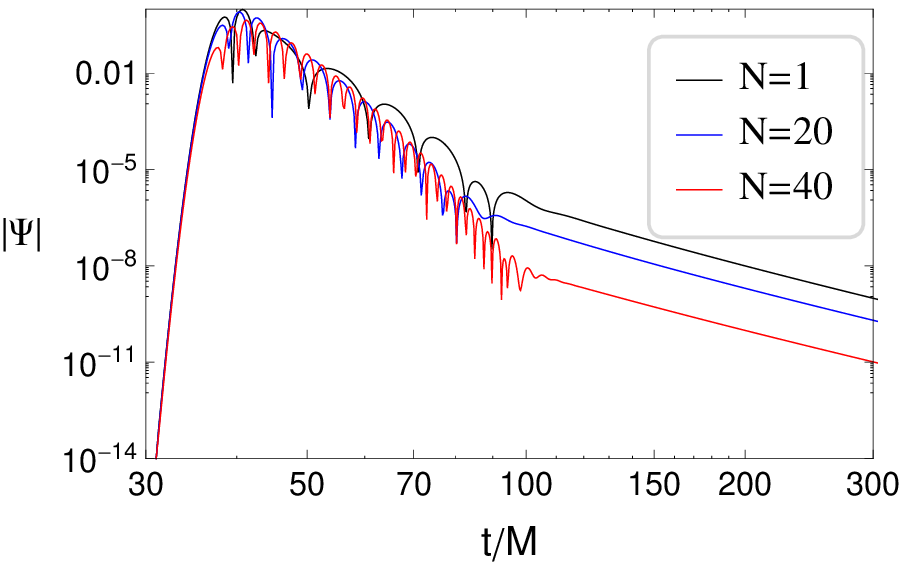}
\includegraphics[width=0.32\linewidth]{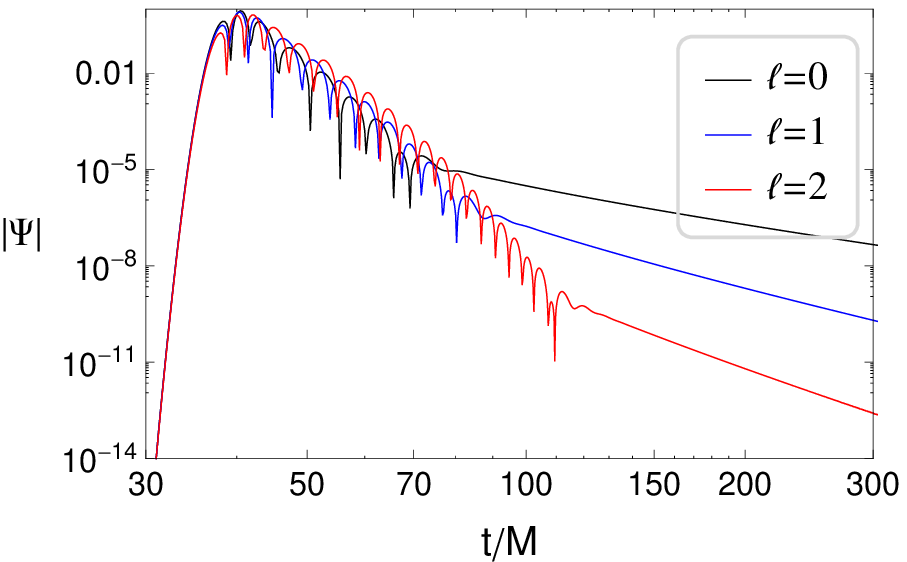}
\end{center}
\caption{\label{fig-time-domain} Evolution of scalar perturbations in the non-singular non-rotating black hole background in conformal gravity as a function of time for different values of the characteristic parameters. Left panel: $\ell=1$, $N=20$. Middle panel: $\ell=1$, $L/M=1.2$. Right panel: $L/M=1.2$, $N=20$. The values of the parameters that are changed are given in legend of each panel.}
\end{figure*}

In order to study the evolution of the scalar perturbations in the non-singular non-rotating black hole spacetimes of conformal gravity, we rewrite the wave equation for the propagation of scalar perturbations~(\ref{wave-eq}) in the so called light cone coordinates
\bear
du=dt-dx, \qquad dv=dt+dx,
\ear
as
\bear\label{time-domain}
\left(4\frac{\partial^2}{\partial u\partial v}+V_s(u,v)\right)\Psi(u,v)=0,
\ear
Thus, Eq.~(\ref{time-domain}) can be integrated numerically. We follow the method that was derived in~\cite{Abdalla:PRD:2006,Rezzolla:CQG:2007} for the discretization of the wave function. Some illustrative time domain profiles of scalar perturbations for different values of the characteristic parameters of the spacetime are given in Fig.~\ref{fig-time-domain}.

The right panel of Fig.~\ref{fig-time-domain} shows that quasinormal modes with the larger value of the multipole number $\ell$ live longer than those with smaller $\ell$, i.e., the imaginary part of the quasinormal frequencies decreases as the value of the multipole number $\ell$ increases.

Fig.~\ref{fig-time-domain} shows that, if we increase the value of the parameter $L$ in the range $L/M<1.2$ ($L/M>1.2$ is excluded by X-ray data~\cite{Bambi:1701.00226}), the damping rate (imaginary part of the quasinormal frequency) of scalar perturbations increases.

Last, we note that, assuming $L/M<1.2$, changing the value of $N$ does not affect much the evolution of the perturbations. On the contrary, for the larger values of $L$ the scalar perturbations lives longer as the value of $N$ increases.

\subsection{Late-time tails and stability}

Fig.~\ref{fig-time-domain} also shows that the quasinormal ringing of the perturbations are followed by the late-time tail, i.e, a power-law falloff of the perturbations. This is typical of any perturbative field in a black hole spacetime. One can see from the left and middle panels of Fig.~\ref{fig-time-domain} that the power-law tails are parallel for different values of the parameters $L$ and $N$ and, therefore, the asymptotical decay laws are independent of the black hole parameters $L$ and $N$. These tails depend only on the multipole number, $\ell$, (the right panel of Fig.~\ref{fig-time-domain}) as typical of perturbations evolution in asymptotically flat spacetimes
\bear
\Psi\sim t^{-2\ell-3},
\ear
Therefore, the power-law tails of the perturbations with higher multipole number, $\ell$, decay faster than those with smaller $\ell$.

One of the main results of our paper is that the spherically symmetric non-singular black hole spacetime in conformal gravity is stable against scalar and electromagnetic perturbations. We came to this conclusion from the facts that the effective potentials for the scalar and electromagnetic perturbations~(\ref{potentials}) are positive definite everywhere of the spacetime and, therefore, the differential operator
\bear
\mathcal{D}=-\frac{\partial^2}{\partial x^2}+V_s(x)
\ear
is a positive self-adjoint operator in the Hilbert space of square integrable functions of $x$. Therefore, all solutions of the perturbative equations of motion with compact support initial conditions are bounded.

\subsection{Quasinormal frequencies}

One of our main goals in this paper is to study the quasinormal modes and the stability of the perturbations in our non-singular non-rotating black hole spacetime in conformal gravity. Here we mainly focus our attention on scalar perturbations, since electromagnetic perturbations do not depend on the conformal factor, with the result that they are the same as those in the Schwarzschild black hole spacetime. The electromagnetic perturbations of Schwarzschild black holes have been already studied by the several authors (see -- Refs.~\cite{Ferrari:PRD:1984,Iyer:PRD:1987,Blome:PLA:1984,Cardoso-Lemos:PRD:2004} and references therein).

To calculate the quasinormal frequencies of scalar perturbations one must solve the second order differential equation~(\ref{wave-eq}) with the potential~(\ref{sc-potential}) in the complex frequency plane, $\omega=\omega_r+i\omega_i$, where the real part of the quasinormal frequency, $\omega_r$, represents the actual oscillations of the perturbation, while the imaginary part of the quasinormal frequency, $\omega_i$, characterizes the dissipation of the perturbations. If we write the perturbations as
\bear
\Psi(t,x)=\psi(x)e^{-i\omega t},
\ear
the wave equation~(\ref{wave-eq}) turns into the time independent form as
\bear\label{wave-eq2}
\left[\frac{\partial^2}{\partial x^2}+\omega^2-V_{s}(r)\right]\psi(x)=0.
\ear
Eq.~(\ref{wave-eq2}) can be solved if we know the boundary condition for the wave at the horizon, $x=-\infty$, and infinity, $x=+\infty$. We impose that the wave at the horizon is purely incoming, and that the wave at spatial infinity is purely outgoing:
\bear
&&\psi(x)\sim e^{-i\omega x} \quad as \quad x\rightarrow-\infty \quad (r\rightarrow2M),\nonumber\\
&&\psi(x)\sim e^{i\omega x} \quad as \quad x\rightarrow+\infty \quad (r\rightarrow+\infty),
\ear
Solving the time independent, second order differential equation~(\ref{wave-eq2}) with the potential~(\ref{sc-potential}) analytically is impossible. Therefore, we use the sixth order WKB method for numerical calculations which is given by the relation
\begin{eqnarray}\label{wkb}
\frac{i(\omega^2-V_0)}{\sqrt{-2V_0''}}+\sum_{j=2}^{6}\Lambda_j=n+\frac{1}{2}
\end{eqnarray}
where a prime (``$\prime$'') denotes a derivative with respect to the tortoise coordinate $x$, and $V_0$ stands for the value of the effective potential at its local maximum, $r=r_0$. The order of the WKB corrections is denoted as $j$, and $\Lambda_j$ is the correction term corresponding to the $j$th order. One can find the expressions of the $\Lambda_j$ terms in~\cite{Iyer87,Konoplya:PRD2003}.

\begin{figure*}[th]
\begin{center}
\includegraphics[width=0.47\linewidth]{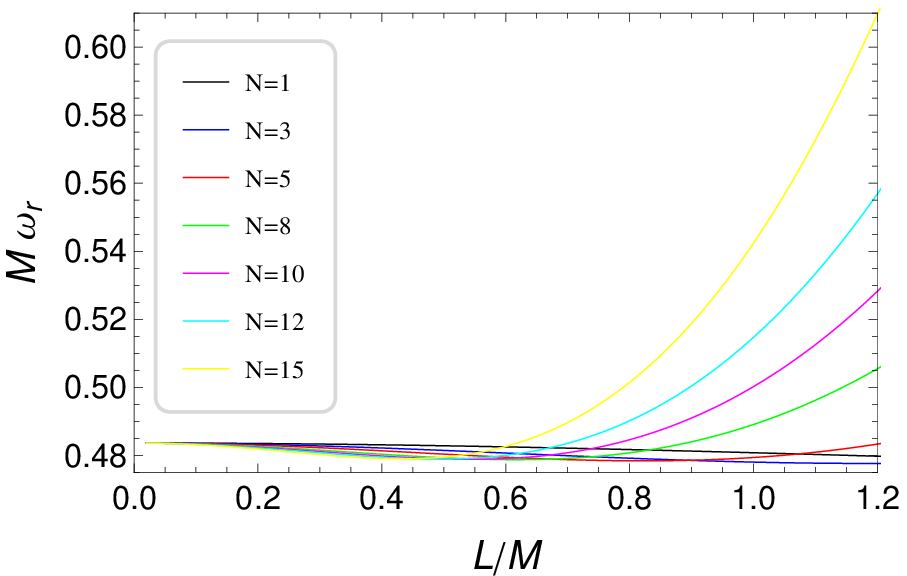}
\includegraphics[width=0.47\linewidth]{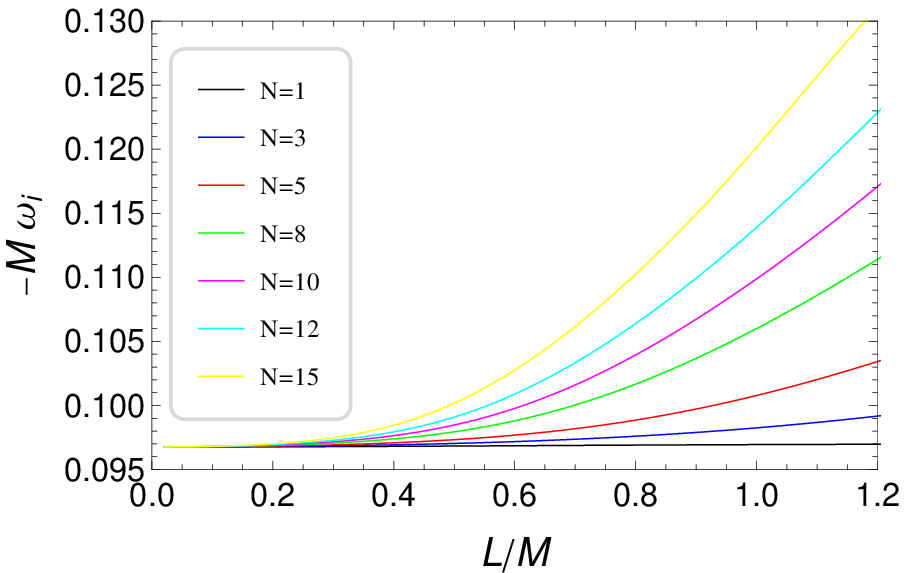}
\end{center}
\caption{\label{fig-scalar} Dependence of the $\ell=2$, $n=0$ quasinormal frequencies of the scalar, $s=0$, perturbation on the parameter $L/M$ for the regular non-rotating black hole in conformal gravity for different values of $N$.}
\end{figure*}

In Fig.~\ref{fig-scalar} and Tabs.~\ref{tab1},~\ref{tab2} and~\ref{tab3}, we show the dependence of the quasinormal frequencies of scalar perturbations on the characteristic parameters.

\begin{figure*}[th]
\begin{center}
\includegraphics[width=0.47\linewidth]{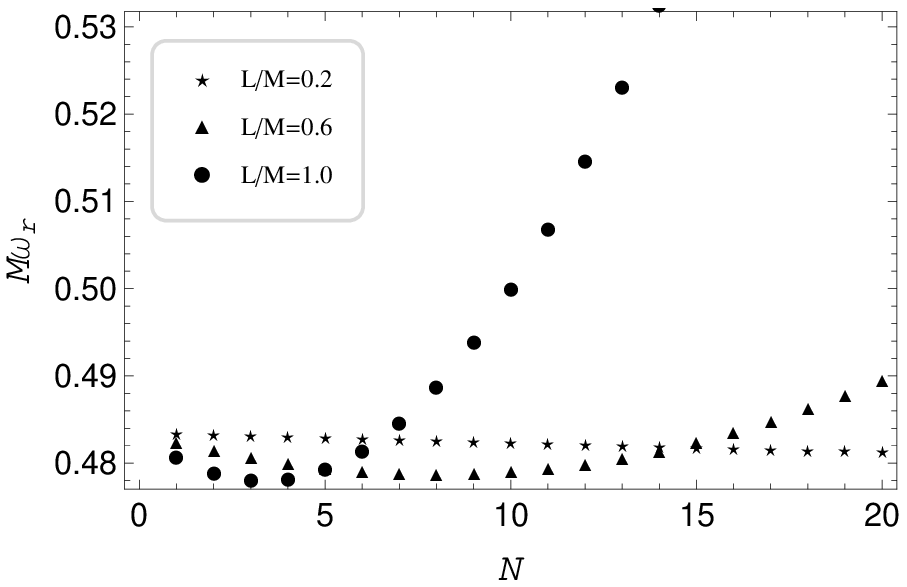}
\includegraphics[width=0.47\linewidth]{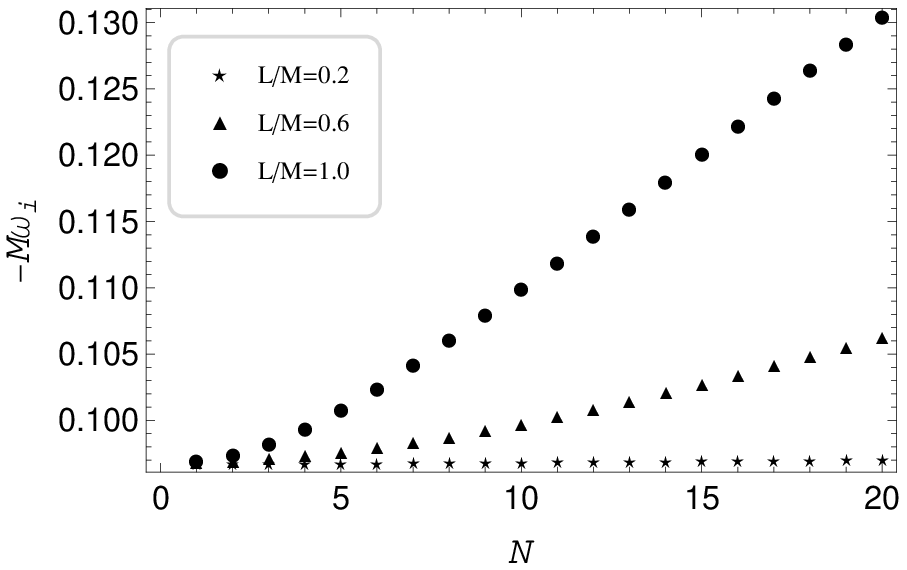}
\end{center}
\caption{\label{fig-dep-n} Dependence of the $\ell=2$, $n=0$ quasinormal frequencies of scalar, $s=0$, perturbations on the degree $N$ for regular non-rotating black holes in conformal gravity for different values of $L$.}
\end{figure*}

\begin{figure}[h]
\centering
\includegraphics[width=0.45\textwidth]{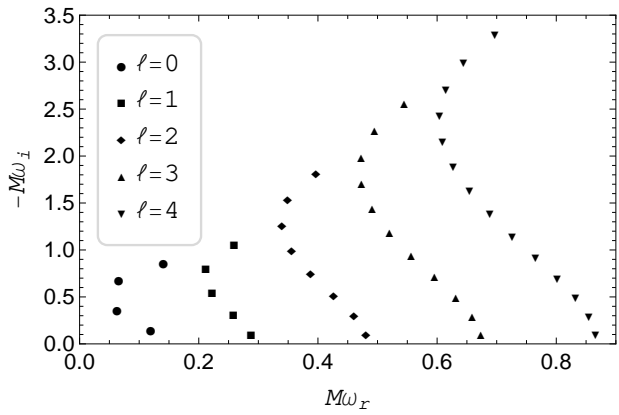}
\caption{\label{fig-non-fundamental} Quasinormal frequencies with $n>0$ for scalar perturbations in non-singular spherically symmetric black holes in conformal gravity with $L/M=1$ and $N=1$ for several values of the multipole number $\ell$. The overtone number, $n$, increases as $n=0,1,2,...$ from bottom to top.}
\end{figure}

In Tab.~\ref{tab1} we present the fundamental quasinormal frequencies (those with the slowest decay rate) of scalar perturbations as a function of the parameter $L$ and for different values of the multipole number $\ell$. $\omega_r$ and $\omega_i$ are obtained by using the sixth order WKB method. However, it is well known that the WKB method has low accuracy in the small $\ell$ regime. Here, to find the quasinormal frequencies for the small values of $\ell$ we used the direct integration method. One can see from Tab.~\ref{tab1} that the real part of the quasinormal frequency, $\omega_r$, decreases as the value of $L$ increases for the small $N$. Note that the difference is small for small values of $L$ if the degree $N$ is small (see the left panel of Fig.~\ref{fig-dep-n}). On the other hand, the magnitude of the imaginary part of the quasinormal frequency, $\omega_i$, remains almost unchanged as the value of $L$ increases when the degree $N$ is small. Therefore, one can say that a small $L$ does not affect much the decay rate of scalar perturbations. The scalar perturbation in the field of the spherically symmetric regular black holes in conformal gravity with small $L$ and $N$ lives shorter on account of the decreased value of the real oscillations and increased value of the damping rate. Moreover, we have also studied the nonfundamental ($n > 0$) QNMs of the scalar perturbations in the field of the nonsingular spherically symmetric black holes in conformal gravity. In Fig.~\ref{fig-non-fundamental} the general features of these nonfundamental QNMs are illustrated.
\begin{table}[h]
\begin{center}
\begin{tabular}{p{0.8cm} p{1.2cm} p{1.5cm} p{1.5cm}}
\hline \hline $\ell$ &$L/M$ & $M\omega_r$ & $M\omega_i$ \\
\hline
0 & 0.0 & 0.110464 & -0.100819 \\
0 & 0.4 & 0.107405 & -0.100335 \\
0 & 0.8 & 0.106088 & -0.087243  \\
0 & 1.2 & 0.056941 & -0.078617 \\
1 & 0.0 & 0.292910 & -0.097762 \\
1 & 0.4 & 0.291945 & -0.097805 \\
1 & 0.8 & 0.289416 & -0.097825 \\
1 & 1.2 & 0.286050 & -0.097665 \\
2 & 0.0 & 0.483642 & -0.096766 \\
2 & 0.4 & 0.483125 & -0.096812 \\
2 & 0.8 & 0.481720 & -0.096910 \\
2 & 1.2 & 0.479769 & -0.096973 \\
\\
\hline
\end{tabular}
\end{center}
\caption{\label{tab1} Fundamental quasinormal frequencies ($n=0$) for scalar perturbations in the spherically symmetric non-singular black hole spacetime in conformal gravity for different values of $\ell$ and $L/M$. Here $N=1$.}
\end{table}

However, in the case of the large values of the degree $N$, the effect of $L$ is significant, see Tab.~\ref{tab2}: $\omega_r$ and $\omega_i$ increase rapidly as $L$ increases in the large values of the degree $N$ (see Fig.~\ref{fig-dep-n}).
\begin{table}[h]
\begin{center}
\begin{tabular}{p{0.8cm} p{1.2cm} p{1.5cm} p{1.5cm}}
\hline \hline $\ell$ &$L/M$ & $M\omega_r$ & $M\omega_i$ \\
\hline
0 & 0.0 & 0.110464 & -0.100819 \\
0 & 0.4 & 0.114925 & -0.140011 \\
0 & 0.8 & 0.241518 & -0.168993  \\
0 & 1.2 & 0.544682 & -0.168520 \\
1 & 0.0 & 0.292910 & -0.097762 \\
1 & 0.4 & 0.281305 & -0.106072 \\
1 & 0.8 & 0.360219 & -0.134909 \\
1 & 1.2 & 0.607217 & -0.156659 \\
2 & 0.0 & 0.483642 & -0.096766 \\
2 & 0.4 & 0.479199 & -0.099492 \\
2 & 0.8 & 0.527633 & -0.116977 \\
2 & 1.2 & 0.717927 & -0.141632 \\
\\
\hline
\end{tabular}
\end{center}
\caption{\label{tab2} As in Tab.~\ref{tab1} but for $N=20$.}
\end{table}
\begin{table}[h]
\begin{center}
\begin{tabular}{p{0.8cm} p{1.2cm} p{1.5cm} p{1.5cm}}
\hline \hline $\ell$ &$N$ & $M\omega_r$ & $M\omega_i$ \\
\hline
0 & 1 & 0.105916 & -0.065272 \\
0 & 5 & 0.116675 & -0.145736 \\
0 & 10 & 0.178803 & -0.160009 \\
0 & 20 & 0.380715 & -0.171040 \\
1 & 1 & 0.287796 & -0.097771 \\
1 & 5 & 0.282387 & -0.110330 \\
1 & 10 & 0.319644 & -0.125070 \\
1 & 20 & 0.464907 & -0.149884 \\
2 & 1 & 0.480789 & -0.096951 \\
2 & 5 & 0.479318 & -0.100797 \\
2 & 10 & 0.499988 & -0.109938 \\
2 & 20 & 0.603374 & -0.130407 \\
\\
\hline
\end{tabular}
\end{center}
\caption{\label{tab3} As in Tab.~\ref{tab1} but for different values of $\ell$ and $N$. Here $L/M=1$ is fixed.}
\end{table}
The imaginary part of the quasinormal frequencies are always negative, $\omega_i<0$. The fact that the value of the wave function $\Psi(t,x)$ is bounded also indicates the correctness of the claim that the spacetime is stable against scalar perturbations.

\subsection{Large multipole number limit}

It is well known that the WKB method is very accurate for the large multipole numbers. Therefore, in order to calculate the quasinormal frequencies in the large multipole number limits analytically we use the first order WKB method. Moreover, it has been shown in~\cite{Konoplya:PRD:2006,Toshmatov:PRD.2015,Toshmatov:PRD.2016} that the quasinormal frequencies are independent of the spin $s$ of the perturbative field. Another method of calculating the QNMs for the large multipole numbers was suggested by Cardoso \textit{et al.}~\cite{Cardoso:PRD:2009}. They have shown that the QNMs in the large multipole number limit are characterized by circular unstable null geodesics and their instability (Lyapunov exponent)~\footnote{In \cite{Konoplya:PLB:2017} it has been shown that these phenomena are violated in the Einstein-Lovelock theory.}. Calculating the QNMs with these two methods we arrive at the same result
\bear\label{large l}
M\omega=\frac{1}{6\sqrt{3}}\left[(2\ell+1)-i(2n+1)\right]+\mathcal{O}\left(\frac{1}{\ell}\right)\ ,
\ear
Interestingly, the QNMs of perturbative fields with arbitrary spin $s$ in the non-singular non-rotating black hole spacetime in conformal gravity do not depend on the conformal factor $S$ and the spin of of the perturbation $s$.

\subsection{Large $N$ limit}

To study the quasinormal frequencies in the large $N$ limit, we use the first order WKB method, i.e., we expand the expression~(\ref{wkb}) in powers of $1/N$, and, taking the first order terms, we obtain the quasinormal frequencies at large values of $N$ as
\bear\label{large N}
\omega_r&\approx&\frac{L^2[4N(r_0-2M)-r_0]+r_0^2(r_0-4M)}{2r_0^2(r_0^2+L^2)},\\
\omega_i&\approx&\left(n+\frac{1}{2}\right)\left[L^4(r_0-6M)(3r_0-8M)+\right. \nonumber\\&&\left.r_0^4(3r_0-10M)(7r_0-16M)-4L^2r_0^2(40M^2+\right. \nonumber\\&&\left.3r_0(r_0-8M))\right]^{1/2}/[r_0^2(r_0^2+L^2)],\nonumber
\ear
In this case the location of the maximum of the effective potential $r_0$ also tends to the finite value
\bear
r_0\approx\frac{1}{9}\left[8M+\frac{64M^2-9L^2}{A^{1/3}}+A^{1/3}\right],
\ear
with
\bear
A&&=512M^3+378L^2M\nonumber\\
&&+9\sqrt{3}L\sqrt{(L^2+4M^2)(3L^2+512M^2)}\ ,\nonumber
\ear
One can see from Eq.~(\ref{large N}) that for large values of $N$, the real part of the quasinormal frequencies, $\omega_r$, is a linear monotonically increasing function of $N$, while the imaginary part, $\omega_i$, does not depend on that and is a linear monotonically increasing function of the overtone number $n$. Interestingly, both parts of the quasinormal mode frequencies do not depend on the multipole number $\ell$.

\section{Conclusion}\label{sec-conclusions}

In the present paper we have studied scalar and electromagnetic perturbations in the family of singularity-free non-rotating black hole spacetimes found in~\cite{Bambi:1611.00865,Modesto:2016max}. We find that scalar perturbations depend on the exact form of the scaling factor $S$, while electromagnetic perturbations in a spherically-symmetric black hole spacetime are not affected by a conformal transformation of the background metric. We have thus focused our study to scalar perturbations, since electromagnetic perturbations are the same as in the Schwarzschild metric and they have been already studied by other authors.

We find that the singularity-free non-rotating black holes found in~\cite{Bambi:1611.00865,Modesto:2016max} are stable under scalar and electromagnetic perturbations (the case of electromagnetic perturbations directly follows from the fact that it is true in the Schwarzschild spacetime). The quasinormal mode spectrum of scalar perturbations depends on the parameters $L$ and $N$. If we assume that the quasinormal mode spectrum for scalar perturbations is not too different from that of gravitational waves\footnote{This assumption was employed in Ref.~\cite{Konoplya:2016pmh} to get model-independent constraints from GW150914, but it has to be taken with some caution. In Einstein's gravity and in some alternative theories of gravity, it is true. In other gravity theories, it is not, and the spectrum for scalar perturbations and gravitational waves is very different. In our case, these metrics are exact solutions in a large class of conformally invariant theories of gravity, and it is possible that in some models it is correct and in other models it is not.}, we can argue that the detection of quasinormal modes from astrophysical black holes can constrain the parameters $L$ and $N$. Note that for $\ell = 2$ (which should be the dominant mode in the case of the coalescence of a binary black hole merging into a single black hole) the impact of $L$ and $N$ can be large and surely detectable. Our calculations cannot be directly applied to the recent observations of gravitational waves by the LIGO experiment because here we have only studied the case of non-rotating black holes. For example, from Tab.~\ref{tab3} we see that for the $\ell = 2$ mode, the difference of the real part of the frequency between the cases $N=1$ and $N=20$ is about 20\%, which is larger than the uncertainties in the measurements of the frequencies of the quasinormal mode observed by LIGO.

It is well known that in the high energy regime the wavelength is almost negligible relative to the horizon scale of the black hole. Therefore, in this regime massless scalar waves follow the null geodesics. Thus, one can consider the classical capture cross section of null geodesics to describe the absorption cross section of massless fields. It is well known from~\cite{Bambi:1611.00865} that the classical capture cross section of null geodesics in the non-singular black hole spacetimes in conformal gravity does not depend on the conformal factor. Therefore, one can say that the conformal factor, i.e., parameters $L$ and $N$, plays no role in the dynamics of the massless scalar waves around regular black holes in conformal gravity in the high energy regime.

\begin{acknowledgments}
B.T., Z.S. and J.S. would like to express their acknowledgments for the institutional support of the Faculty of Philosophy and Science of the Silesian University in Opava, the internal student grant of the Silesian University (Grant No. SGS/14/2016) and the Albert Einstein Centre for Gravitation and Astrophysics under the Czech Science Foundation (Grant No.~14-37086G). The work of C.B. was supported by the NSFC (Grant No.~U1531117), Fudan University (Grant No.~IDH1512060), and the Alexander von Humboldt Foundation. The research of B. A. is supported in part by Grant No.~VA-FA-F-2-008 of the Uzbekistan Agency for Science and Technology, and by the Abdus Salam International Centre for Theoretical Physics through Grant No.~OEA-NT-01 and by the Volkswagen Stiftung, Grant No.~86 866. This research is partially supported by Erasmus + exchange grant between Silesian University in Opava and National University of Uzbekistan.
\end{acknowledgments}

\newpage

\label{lastpage}

\end{document}